# Emergence of transverse spin in optical modes of semiconductor nanowires

**M.H. Alizadeh**[1,2] **and Björn M. Reinhard**[1,2]

[1]*Photonics Center, Boston University, 8 Saint Mary's Street, 02215, USA*
[2]*Chemistry Department, Boston University, 590 Commonwealth Avenue, 02215, USA*
[*]*halizade@bu.edu, bmr@bu.edu*

**Abstract:** The transverse spin angular momentum of light has recently received tremendous attention as it adds a new degree of freedom for controlling light-matter interactions. In this work we demonstrate the generation of transverse spin angular momentum by the weakly-guided mode of semiconductor nanowires. The evanescent field of these modes in combination with the transversality condition rigorously accounts for the occurrence of transverse spin angular momentum. The intriguing and nontrivial spin properties of optical modes in semiconductor nanowires are of high interest for a broad range of new applications including chiral optical trapping, quantum information processing, and nanophotonic circuitry.

## 1. Introduction

The unusual spin angular momentum (SAM) and spin momentum of evanescent and structured electromagnetic (EM) fields have recently drawn considerable attention [1-18]. The new SAM is considered "unusual" because it is transverse to the direction of the propagation. In the prevalent picture of EM spin, the vectorial nature of the EM fields leads to the inherent polarization degree of freedom. Depending on its circular handedness, a photon can be of spin +1 or -1. In the quantum picture, due to the masslessness of photons, the concepts of spin and helicity cease to differ and can be used interchangeably. SAM is not the only type of angular momentum that light can possess. The angular momentum of light comprises an orbital and a spin part. While SAM originates from spinning electric and magnetic fields, orbital angular momentum (OAM) stems from the phase gradient of the wavefront [14, 15]. This distinction is not fundamental and in fact, spin-to-orbital angular momentum conversion may occur in both inhomogeneous anisotropic media and in tightly focused beams [19-22]. This interchangeability, sometimes called spin orbit interactions, can lead to novel optical phenomena that appear in length scales compared to the wavelength of light and in which, SAM can control light's spatial intensity distribution and propagation path [20]. SAM is a measure of the circular polarization of EM fields and for propagating waves is either parallel or anti-parallel to the direction of the propagation. It can be written as [9]:

$$S = S_e + S_m, \quad S_e = \frac{\varepsilon_0}{4\omega}\mathrm{Im}\{E \times E^*\}, S_m = \frac{\mu_0}{4\omega}\mathrm{Im}\{H \times H^*\} \qquad (1)$$

One readily observes that for a propagating linearly polarized light, SAM goes to zero and only for EM fields with non-zero ellipticity is nontrivial. For propagating circularly polarized light (CPL), in particular, SAM is non-zero and is either parallel or anti-parallel to the wavevector. For evanescent and non-paraxial EM fields, however, a new type of SAM can emerge. Recently it was shown that unlike paraxial beams that carry strictly longitudinal spin, evanescent waves possess new type of spin which is transverse to the direction of propagation

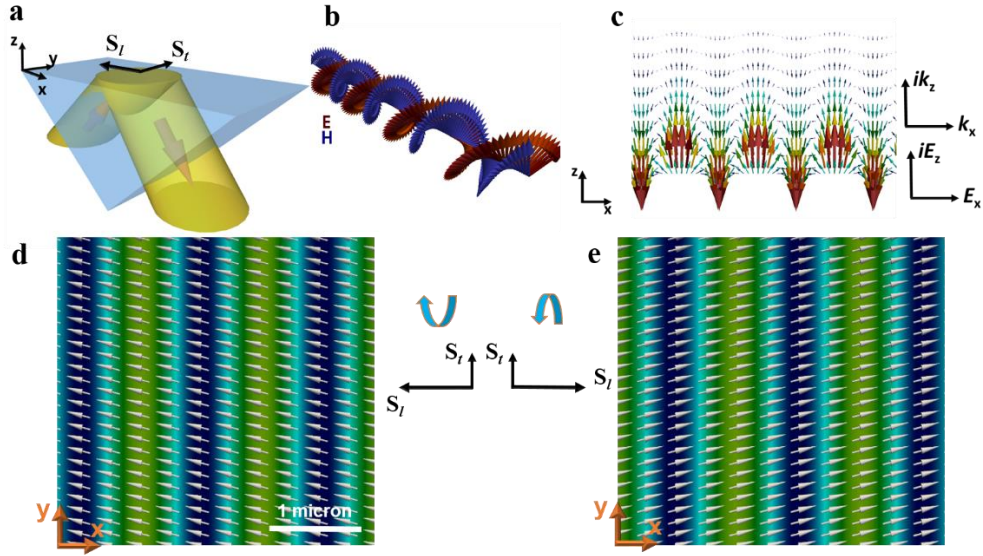

Fig.1. (a) Schematics of exciting an evanescent wave in glass-air interface by Total Internal Reflection. (b) The vectorial behavior of the **E** and **H** of the evanescent wave excited by a circularly polarized beam. It is observed that the evanescent wave preserves the longitudinal spin of the excitation beam (c) Vectorial electric field in the plane of decay, xz plane. The evanescence of the field combined with the transversality condition leads to spinning electric field in xz plane, which in turn, results in the transverse SAM. (d), (e) The SAM is depicted in a background of electric field distribution. For a circularly polarized evanescent plane wave, the total SAM is a sum of helicity-dependent longitudinal SAM and the helicity-independent transverse SAM. When the circular polarization of the excitation beam reverses, the longitudinal SAM reverses direction while the transverse SAM does not.

[2]. Consider an evanescent wave propagating in x direction and decaying in z direction (See Fig. 1(c)). For a surface without free charges and currents $\nabla \cdot \boldsymbol{E} = \boldsymbol{0}$, which leads to $\boldsymbol{k} \cdot \boldsymbol{E} = \boldsymbol{0}$. This is known as the transversality condition [4]. The evanescence of the wave indicates that the wavevector has an imaginary component is the decaying direction: $\boldsymbol{k} = k_x \hat{x} + i k_z \hat{z}$, which can be combined with the transversality condition to deduce the existence of an imaginary longitudinal electric field, such that, $E_x = -i(k_z/k_x)E_z$. This phase-shifted longitudinal component dictates the rotation of the electric field of the evanescent wave in the plane of propagation, which leads to SAM in a direction transverse to the direction of propagation. It can be shown that the transverse SAM can be written as $S_t \propto \Re(\boldsymbol{k}) \times \Im(\boldsymbol{k})$, which has the profound consequence that this novel spin, unlike the longitudinal spin, does not depend on the polarization of the wave and is locked to the wavevector [2]. This is illustrated in Fig.1, where the distributions for spin density of an evanescent field under circularly polarized excitation at the glass-air interface is shown. In Fig. 1(c) and 1(d), the longitudinal spin of the evanescent wave is reversed by changing the circular polarization of the excitation beam from left to right, and as a result, its direction changes from parallel to anti-parallel with the wavevector. The transverse component of the spin, however, does not change the direction upon polarization inversion of the excitation beam. So far, the emergence of transverse spin has been studied for evanescent waves in planar interfaces [4], for surface plasmon polaritons [2, 9] and for non-paraxial beams [8, 9, 23]. Lateral chiral optical forces that result from transverse SAM are among the major motivations behind this surge in interest in transverse SAM. Such lateral chiral forces appear in opposite directions for chiral objects of opposite handedness [10, 24-39]. Extending the concept of transverse SAM to photonic cavity modes, however, poses a challenge, as most of such modes are confined to the photonic resonators, be it an optical fiber, a micro-toroid or other photonic cavities. Here we demonstrate the emergence of strong transverse SAM in weakly-guided optical modes of a semiconductor nanowire. These weakly guided modes are photonic modes that, despite being bound to the semiconductor nanowire, have significant modal "spill-over" to the surrounding of the photonic resonator [40]. We show that the vectorial behavior of the evanescent EM fields of a weakly guided mode, which itself is a consequence of transversality condition, results in emergence of SAM in a direction transverse to the propagation of the optical mode. To this end, through scattering cross section calculations, we first investigate the conditions for excitation of the weakly-guided mode in vertical nanowires. In the next step, we analyze the vectorial behavior of the EM fields and demonstrate the emergence of SAM in these

modes.

## 2. Weakly guided modes in vertical nanowires

The electromagnetic modes of an infinitely long nanowire with radius R can be found by solving the dispersion relation [41-46]:

$$\left[\frac{\mu_c}{k_c R}\frac{J'_m(k_c R)}{J_m(k_c R)} - \frac{\mu}{kR}\frac{H'_m(k_c R)}{H_m(k_c R)}\right] \times \left[\frac{\varepsilon_c}{k_c R}\frac{J'_m(k_c R)}{J_m(k_c R)} - \frac{\varepsilon}{kR}\frac{H'_m(k_c R)}{H_m(k_c R)}\right] = m^2 \frac{(k_z R)^2}{(\omega R/c)^2}\left[\frac{1}{(kR)^2} - \frac{1}{(k_c R)^2}\right]^2 \qquad (2)$$

Where m is an integer, $J_m$ and $H_m$ are cylindrical Bessel and Hankel functions, respectively. $k_c$ and $k$ are the components of the wavevector of the modes which are transverse to the cylinder long axis along which the modes propagate. $\mu_c$ and $\mu$ are the permeabilities of the cylinder and the medium, respectively, and $\varepsilon_c$ and $\varepsilon$ are the permittivities of the cylinder and medium respectively. Depending on the nature of the transverse wavevector of a mode, it can be categorized as either leaky or guided. A purely imaginary transverse wavevector results in a

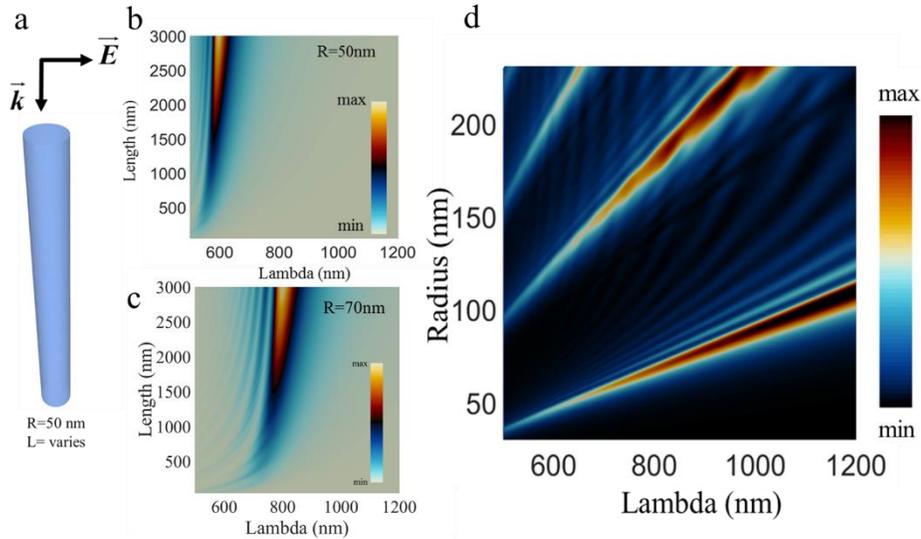

Fig.2 (a) The schematics of excitation of the weakly-guided mode. (b) For a vertical nanowire with the length of 3 μm and the radius of 50 nm, only the fundamental weakly-guided mode, $HE_{11}$, is excited. (c) When the radius is increased to 70 nm, still only the $HE_{11}$ mode can be excited. This is because the radial size of the nanowire is too small to sustain higher modes. (d) This is further confirmed when the scattering cross section is calculated for a group of nanowires with the length of 3 μm but with different radii. It is seen that for radii smaller than about 100 nm, only the fundamental mode is excited. For larger radii, however, higher modes can be excited.

real wavevector along the cylinder axis, hence a propagating mode along it. A real transverse wavevector, however, leads to modes which leak out of nanowire. When m=0, the transverse magnetic and transverse electric modes can independently exist with radial mode distributions of different orders, namely for different values of $l$ in $TE_{0l}$ and $TM_{0l}$ modes. For the case of nanowires with finite length, however, an analytical solution does not exist and the problem should be tackled numerically. For our case, we performed full-wave EM simulations to obtain scattering cross section of a free standing silicon nanowire with the radius of 50 nm, excited under normal incident linearly polarized light. The schematics of the excitation geometry is shown in Fig. 2(a). We studied nanowire lengths from 50 nm to 3000 nm for a constant radius of 50 nm. The motivation for studying such a vast range of heights is to gain insight into behavior of the optical modes when the shape of the resonator shifts from a disk to a nanowire. As is seen in Fig. 2(b), only one mode is efficiently excited in the structure at around λ=600 nm, which occurs only for heights larger than a threshold. The scattering cross section spectrum is dominated by the fundamental waveguide mode of the nanowire, which can be excited only when the height of the structure is long enough to support such a propagating mode and the momentum-matching condition can be met. Also, it is evident that the cross section of the structure is too small to sustain higher modes with larger radial and azimuthal modal numbers. The linear polarization along with the angle of incidence of the excitation beam breaks the azimuthal symmetry and as a consequence the incoming wave cannot couple to any $TE_{0l}$ and $TM_{0l}$ modes, which inherently possess axially symmetric modal structures. polarized in x direction and couples to the fundamental $HE_{11}$ mode, which propagates with real $k_z$ in positive z direction and is evanescent with $ik_x$ in x direction. As expected, the mode has maxima of its electric field intensity in the direction of the polarization of the incident light and has mode localization mainly outside the nanowire with decaying intensity due to its evanescence. (See Fig.3 (c)). The existence of an imaginary $k_x$ combined with the transversality condition, ***k.E = 0***, indicates a π/2 phase-shifted electric field component in z direction. This phase shift between $E_x$ and $E_z$ induces a nontrivial ellipticity of weakly-guided mode in xz plane. This can be clearly seen in Fig. 3(d), where the vectorial electric field is plotted for $\lambda = 600$ nm, at which wavelength, the optimum coupling of the incoming light to the $HE_{11}$ mode occurs. The electric field, evidently, exhibits a distinct rotational behavior and nontrivial ellipticity, which in turn, should generate unusual transverse SAM directed normal to the plane of rotation. What makes this transverse SAM more intriguing is the lack of any spin feature in the excitation light. It should be noted that the emergence of transverse spin is contingent upon excitation of a waveguide mode with an evanescent tail. When the incoming light does not couple to such a mode, the ellipticity of the mode is trivial. This is seen in Fig. 3(e), where similar to Fig. 3(d), the vectorial behavior of the electric field is shown. In this case, however, it is plotted for $\lambda = 900$ nm, at which wavelength, almost no appreciable coupling occurs between the incoming light and an optical mode, as can be verified from the scattering cross section plot, Fig. 2(b). The incoming light, without exciting any waveguide mode, travels along the nanowire without "seeing it", thus its electric field exhibits the vectorial behavior of a plane wave with no ellipticity. As was mentioned previously, the presence of fields with nontrivial ellipticity signals the emergence of transverse spin in the optical mode. The simulated spin density distribution on a surface 10 nm above the nanowire, is shown in Fig. 4(a) and its vectorial behavior in Fig. 4(c). A closer look at Fig. 4(a) reveals that, as expected, the maxima of SAM coincide with spots with higher ellipticity. Also, the spin density has larger magnitude in yz plane than in xz plane. In order to explain this point, we should note that HE modes, unlike EH modes, are dominantly magnetic and the magnetic intensity maxima occur at the plane normal to xz, i.e. yz plane. This is illustrated in Fig. 3(b) where it is observed that the magnetic field peaks in the perpendicular plane to xz, i.e. yz plane and in Fig. 3(b) where the vectorial magnetic field of the mode, propagating right above the nanowire surface, is shown for yz plane. It is clearly seen that the magnetic field exhibits a strong ellipticity on yz plane. The ellipticity of the magnetic field in the xz plane as well as that of the electric field in the yz plane are trivial (not shown here). The sum of magnetic and electric spin density results in a quasi-symmetrical azimuthal distribution of spin with larger magnitude in y direction. Fig. 4(c), is a bird-eye view of the vectorial SAM and it clearly confirms the above analysis, with the spin showing an azimuthal distribution with larger magnitude in y direction. The emergence of this transverse spin in the vicinity of the nanowire is of fundamental importance, as it adds a whole new degree of freedom to explore. It is interesting that an optical mode excited by an incoming wave with no spin features, not only possesses spin, but its spin is normal to its propagation direction. It should be noted, however, that the azimuthal symmetry of the spin combined with its vectorial nature makes its volume integral zero, which is expected from conservation of angular momentum. It should be mentioned that what makes the weakly-guided mode special, in the context of transverse SAM, is its significant modal "spill-over", which is essential in achieving large and accessible transverse SAM. This feature lacks in higher optical modes of nanowires. As the diameter of the nanowire grows, it becomes capable of sustaining optical modes with higher radial and azimuthal numbers. Such higher modes are mostly confined to the resonator and do not show as much of a significant modal "spill-over" as the weakly-guided mode. To illustrate this point, the electric field and SAM for a silicon nanowire with the same height of 3 μm, but with a radius of 150 nm, were calculated at λ=640 nm which corresponds to a higher mode of the nanowire. The electric field distribution at the nanowire cross section and its vectorial behavior are shown in Fig. 5(a) and Fig. 5(c), respectively. The corresponding SAM is illustrated in Fig. 5(b). The excitation of a higher mode is evident from the radial and azimuthal modal distribution with higher number of nodes and peaks at the displayed cross section. The most striking difference from the weakly-guided mode, however, is the degree of the confinement of the mode. As can be seen in Fig. 5(a), the electric field is mostly confined in the nanowire, showing small modal delocalization to the surrounding.

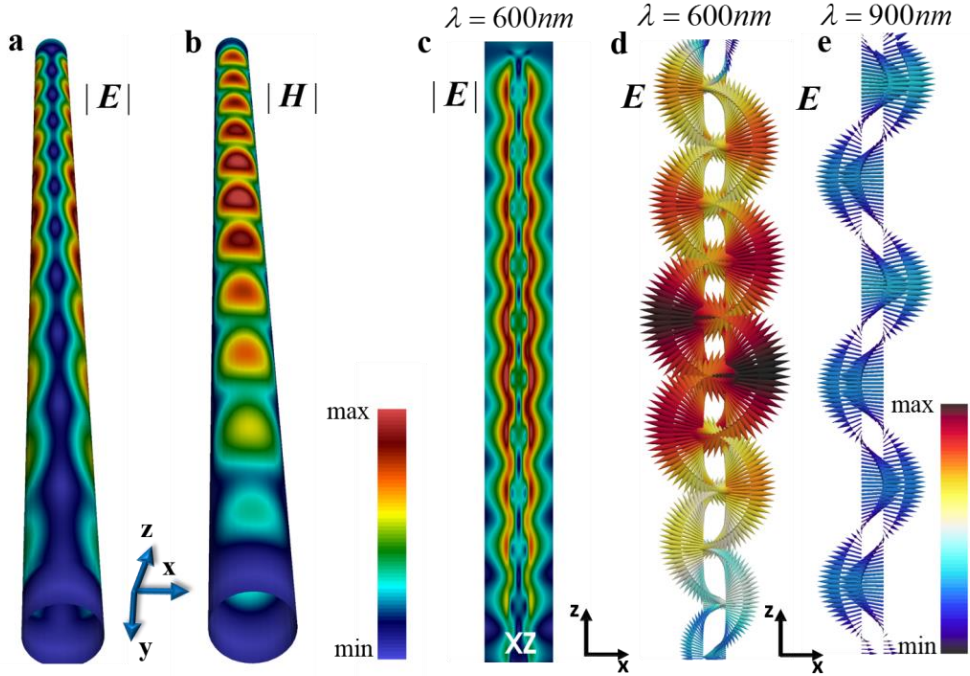

Fig.3. (a) The electric field distribution is shown on a cylindrical surface, 10 nm above the nanowire surface for λ=600 nm. As expected, the peak of the electric field lies along the polarization axis of the incident light (b) Similar to (a), but in this case for the magnetic field. The peaks occur perpendicular to the polarization plane, i.e. in yz plane. (c) The distribution of the electric field is shown in the xz plane, passing through the center of the nanowire. The most relevant feature of the $HE_{11}$ mode, which is its significant modal spill-over to the surrounding, is clearly seen in this plot. The mode travels along the nanowire with a large evanescent tail. (d) The spinning electric field at the modal resonance results in SAM transverse to the direction of the mode propagation. (e) When the incident light does not couple to any modes, it travels along the nanowire like a plane wave.

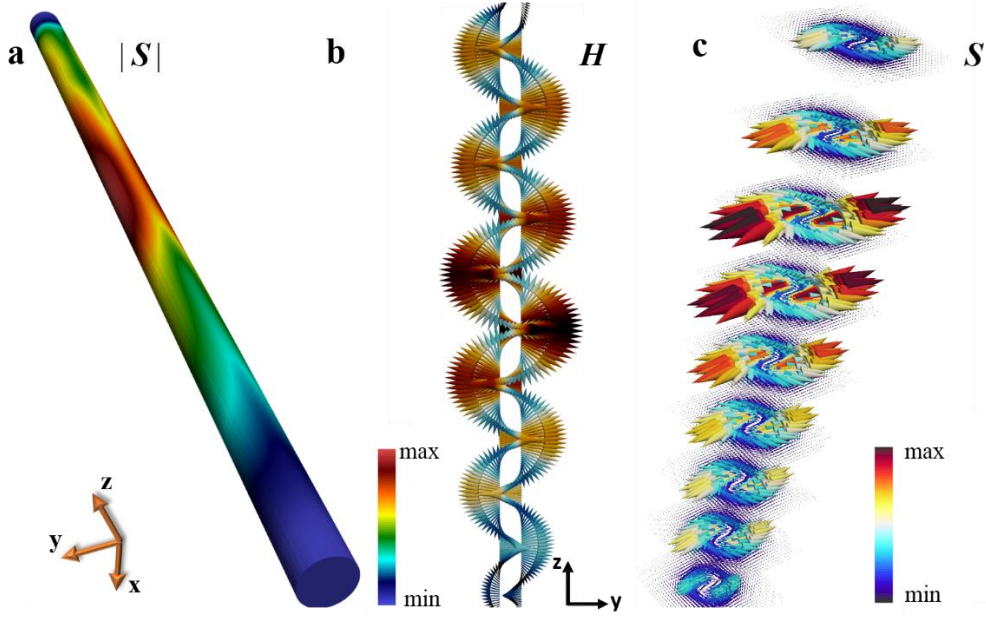

Fig.4. (a) A bird's eye view of SAM distribution is shown on a cylindrical surface, 10 nm above the nanowire surface for λ=600 nm. Interestingly the spin density has larger magnitude in y direction than in x direction. The reason is the occurrence of the peaks of magnetic field in yz plane. This is further confirmed when we look at the vectorial magnetic field on yz plane. (b) Vectorial magnetic field on yz plane for λ=600 nm. The spinning magnetic field has the major contribution to the transverse spin on this plane. (c) Full vectorial SAM shown for λ=600 nm. Transverse spin lies along azimuthal direction which is transverse to the **k** of the mode which is in z direction. It shows azimuthal distribution with an asymmetry caused by larger value in y direction (redder vectors).

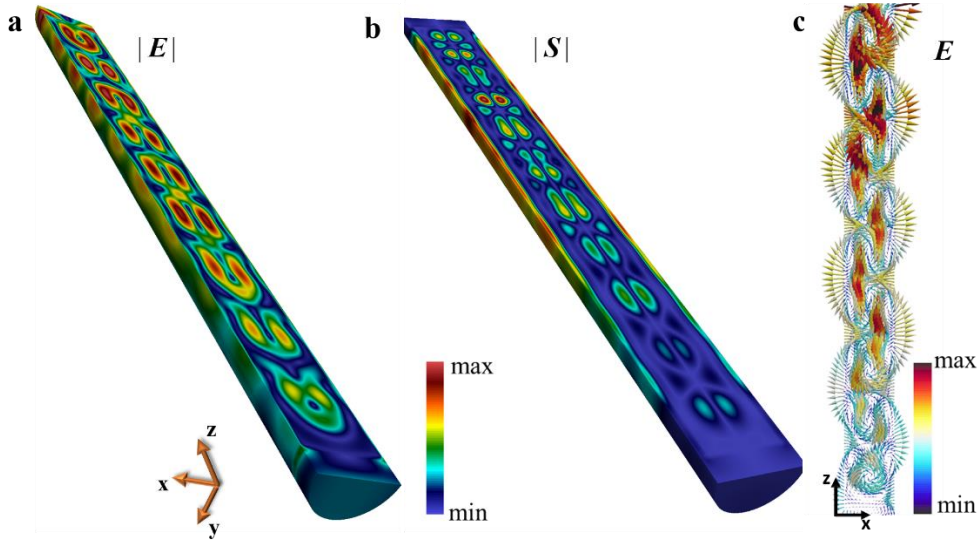

Fig.5. (a) Electric field distribution for a nanowire with a radius of 150 nm and length of 3μm shown for λ=640 nm. This nanowire is radially large enough to support higher modes, which are mostly confined to the structure. (b) As a consequence of the confinement of the mode, SAM appears mainly within the structure. (c) A closer look at the vectorial electric field confirms its confinement within the nanowire and its small evanescent tail.

This is more appreciable in Fig. 5(c), where one can see that the stronger electric fields (redder vector fields) are largely confined inside the nanowire. A small evanescent tail, which is inevitable in subwavelength nanowires, remains and results in weak ellipticity in the vicinity of the structure.

## 3. Conclusion

To conclude we demonstrated that the weakly guided mode of a subwavelength nanowire possesses transverse SAM. We studied the nature of this mode and showed that an intrinsic field delocalization into the ambient medium bestows it with a strong evanescent electromagnetic tail. We then demonstrated that this evanescence combined with the transversality condition, which itself is a direct consequence of Maxwell's equations, accounts for the transverse SAM in the vicinity of these photonic structures. This finding is of great importance as it paves the path to new transverse SAM based applications. One immediate application may be the generation of chiral lateral optical forces that can act on chiral objects and achieve an enantiomer selective optical trapping. Also, in another frontier, transverse SAM of optical modes of semiconductor nanowires may find applications as a new degree of freedom in quantum information processing and data storage. In particular, CMOS-compatibility of semiconductor nanowires adds to their potential applicability in this regard.

**Acknowledgments**

This work was supported by the U.S. Department of Energy, Office of Basic Energy Sciences, Division of Materials Science and Engineering under Award DOE DE-SC0010679.